\documentclass[%
 aip,
 rsi,% Review of Scientific Instruments
 amsmath,amssymb,
 reprint,%
]{revtex4-2}

\usepackage{graphicx}% Include figure files
\usepackage{dcolumn}% Align table columns on decimal point
\usepackage{bm}% bold math
\newcommand{\Eq}[1]{Eq.\,\ref{#1}}% \Eq{abc}
\newcommand{\Fig}[1]{Fig.\,\ref{#1}}% \Fig{abc}
\newcommand{\Sec}[1]{Sec.\,\ref{#1}}% \Sec{abc}
\newcommand{\be}{\begin{equation}}
	\newcommand{\ee}{\end{equation}}
\newcommand{\bea}{\begin{eqnarray}}
	\newcommand{\eea}{\end{eqnarray}}
\newcommand{\bd}{\begin{displaymath}}
	\newcommand{\ed}{\end{displaymath}}
\newcommand{\ben}{\begin{enumerate}}
	\newcommand{\een}{\end{enumerate}}

\newcommand{\Onlinecite}[1]{Ref.\,\onlinecite{#1}} % \Onlinecite{abc}
\newcommand{\fss}{\mathrm{fs^{2}}} %% fs^2 for GDD

\mathchardef\mhyphen="2D %% a proper hyphen for math mode

\newcommand{\rcm}{\ensuremath{\mathrm{cm}^{-1}}} %% omega Pump

\newcommand{\Ip}{\ensuremath{I_\mathrm{p}}} %% pump intensity
\newcommand{\Is}{\ensuremath{I_\mathrm{s}}} %% Stokes intensity
\newcommand{\Si}{\ensuremath{S_\mathrm{int}}} %% Edge size
\newcommand{\Ng}{\ensuremath{N_\mathrm{g}}} %% number of galvo steps in each direction
\newcommand{\Av}{\ensuremath{\bar{A}_\mathrm{v}}} %% Average pixel value in valid tiles

\newcommand{\pcij}{\ensuremath{\tilde{p}_{ij}}} %% corrected pixel value
\newcommand{\pctij}{\ensuremath{\breve{p}_{ij}}} %% tiling corrected pixel value
\newcommand{\rnm}{\ensuremath{\mathbf{r}_{nm}}} %% focal point position in pixels at reference galvo position
\newcommand{\rp}{\ensuremath{\bar{\mathbf{r}}}} %% polynomial fit to focal point position at reference galvo poisition
\newcommand{\Sg}{\ensuremath{S_\mathrm{g}}} %% grid spacing on camera image in pixels 
\newcommand{\sigr}{\ensuremath{\sigma_\mathrm{r}}} %% read noise
\newcommand{\sigrs}{\ensuremath{\sigma_\mathrm{s}}} %% summed read noise

\newcommand{\nuf}{\ensuremath{\nu_\mathrm{f}}} %% frame rate
\newcommand{\texp}{\ensuremath{\tau_\mathrm{e}}} %% exposure time
\newcommand{\tdel}{\ensuremath{\tau_\mathrm{d}}} %% delay time between exposure end and trigger
\newcommand{\cgain}{\ensuremath{g_\mathrm{c}}} %% camera gain

\begin{document}

%\preprint{AIP/123-QED}

%\title[Sample title]{Sample Title:\\with Forced Linebreak\footnote{Error!}}% Force line breaks with \\
%\thanks{Footnote to title of article.}
\title[Thousand foci CARS]{Thousand foci coherent anti-Stokes Raman scattering microscopy}

\author{Dominykas Gudavičius}
\affiliation{ 
	Light Conversion, Keramiku st. 2B, LT-10233 Vilnius, Lithuania%\\This line break forced with \textbackslash\textbackslash
}%
\affiliation{ 
	School of Physics and Astronomy, Cardiff University, The Parade, Cardiff CF24 3AA, United Kingdom%\\This line break forced with \textbackslash\textbackslash
}%

\author{Lukas Kontenis}%
\affiliation{ 
	Light Conversion, Keramiku st. 2B, LT-10233 Vilnius, Lithuania%\\This line break forced with \textbackslash\textbackslash
}% \textbackslash\textbackslash

\author{Wolfgang Langbein}%
 \email{langbeinww@cardiff.ac.uk}
\affiliation{ 
School of Physics and Astronomy, Cardiff University, The Parade, Cardiff CF24 3AA, United Kingdom%\\This line break forced with \textbackslash\textbackslash
}%

\date{\today}% It is always \today, today,

\begin{abstract}
We demonstrate coherent anti-Stokes Raman scattering (CARS) microscopy with 1089 foci, enabled by a high repetition rate amplified oscillator and optical parametric amplifier. We employ a camera as multichannel detector to acquire and separate the signals from the foci, rather than using the camera image itself. This allows to retain the insensitivity of the imaging to sample scattering afforded by the non-linear excitation point-spread function, which is the hallmark of point-scanning techniques. We show frame rates of 0.3\,Hz for a megapixel CARS image, limited by the camera used. The laser source and corresponding CARS signal allows for at least 1000 times higher speed, and using faster cameras would allow acquiring at that speed, opening a perspective to megapixel CARS imaging with more than 100\,Hz frame rate.
\end{abstract}

%Use showkeys class option if keyword display desired
\maketitle

\section{\label{sec:Intro}Introduction}

Coherent Raman scattering (CRS) allows microscopy with chemical selectivity using the vibrational response of the chemical components of the sample without adding fluorescent dyes.\cite{ZumbuschPLR13,ZhangARBE15,ZhangJPP21,LinEL23} Driving the vibrations by an optical excitation which is intensity modulated at the vibrational frequency, the vibrations are in turn modulating the polarisability, creating coherent side-bands of the optical excitation fields. These sidebands are the CRS, and their amplitude and phase encode the vibrational spectrum of the medium probed.
Since its revival about 2 decades ago\cite{ZumbuschPRL99} enabled by the development of suited commercial laser sources, the field has developed a wide range of methods of excitation and detection, with different advantages and drawbacks. \cite{ZumbuschPLR13,LinEL23}

The most common scheme uses two synchronised optical pulses, the pump pulse and the longer wavelength Stokes pulse, with a frequency difference given by the vibrational frequency of interest. To achieve high intensities while keeping the average power below sample damage, pulses of picosecond duration, comparable to the typical vibrational coherence times in condensed matter, with repetition rates in the 1-100\,MHz range are employed, having a duty cycle in the $10^{-4}$ to $10^{-6}$ range. Focussing these pulses to the diffraction limit of high numerical aperture (NA) objectives of about 0.5\,\textmu m size enables achieving peak intensities of the order of $10^{11}-10^{12}$\,W/cm$^{2}$ for average powers of tens of milliwatts, suitable for efficient CRS generation just below typical saturation levels\cite{GongNPho20} around $10^{13}$\,W/cm$^{2}$. Scanning this focus across the sample is then used to provide an image, similar to other multiphoton techniques recording two-photon fluorescence (TPF), second harmonic generation (SHG), or third harmonic generation (THG). The non-linearity of the CRS signal with the excitation intensity enables optical sectioning without imaging in the detection, an advantage compared to confocal or other structured illumination microscopes, specifically for inhomogeneous samples exhibiting significant diffuse scattering.

The high frequency sideband of the pump, called coherent anti-Stokes Raman scattering (CARS), and the low frequency sideband of the Stokes, call coherent Stokes Raman scattering (CSRS), can be detected free of excitation background, while the high frequency sideband of the Stokes interferes with the pump, called stimulated Raman loss (SRL), and the low frequency sideband of the pump interferes with the Stokes, and is called stimulated Raman gain (SRG). Notably, sidebands are also created by purely electronic non-linearities such as the Kerr effect, as well as by vibrational resonances at higher frequencies, providing a background to the resonant response.
These non-resonant responses are in quadrature to the pump or Stokes beams in SRL and SRG, respectively, yielding in lowest order only a change of phase without changing the amplitude. Thus by detecting the transmitted power, which is insensitive to the phase, only resonant responses are measured, rendering SRL and SRG spectra similar to Raman spectra. However, the relative changes are typically very small, in the $10^{-4}$ to $10^{-7}$ range, so that high-frequency modulation schemes and shot-noise limited laser sources are required.  

Detecting the CARS or CSRS power, free of excitation background, does not have these requirements, but the signal contains the interference between resonant and non-resonant response, creating asymmetric resonance lineshapes, complicating the interpretation of the contrast. To retrieve a signal linear in the concentration of the chemical components, the complex susceptibility can be determined by phase retrieval using the causality of the response and the measured dependence on the vibrational frequency.\cite{VartiainenOE06,LiuOL09,MasiaAC13} 

The point scanning approach has been optimised and is now limited by the maximum signal available due to signal saturation by excitation of higher vibrational levels \cite{GongNPho20} or heating, and allows for video rate imaging.\cite{EvansPNAS05,EvansARAC08,SaarS10} However, small absorbing regions can lead to sample damage at the high intensities and repetition rates used, both by heating and by avalanche breakdown. Recently there was some effort to use squeezed light \cite{XuOL22} to reduce the shot-noise in SRL and SRG, with modest success (less than a factor of two in signal to noise) limited by the un-squeezing of the detected light by losses due to scattering in the sample and finite transmission of optical elements.  Thus the only clear way forward to higher speed is to distribute the excitation into many focal spots, or use a wide-field illumination. This requires suited laser sources providing the larger pulse energy required to provide an intensity sufficiently high for efficient CRS generation across the extended excitation region. To limit the average power on the sample, the repetition rate of the laser pulses need to be reduced, which also allows the laser-induced heating and electronic excitations to dissipate between pulses, increasing damage thresholds.

Early wide-field approaches used nanosecond pulsed lasers with repetition rates in the 1-100\,Hz range and a non-colinear geometry for phase matching, enabling to choose between dark-field (phase-mismatched) and bright-field (phase-matched) imaging.\cite{HeinrichAPL04} The method was applied to image 2D cell cultures,\cite{HeinrichOE08} but the coherent wide-field nature of the signal created significant interference artefacts, specifically relevant for thicker and more scattering samples, and furthermore did not provide sectioning. This drawback was addressed using random speckle illumination, which after speckle averaging  resembles a spatially incoherent illumination and thus removes the interference artefacts.\cite{HeinrichNJP08} Multimode optical fibres were used to generate the speckles, with the Stokes beam of a coherence length much larger than the phase delay distribution across the fibre modes retaining the spatial speckle pattern, while the pump beam of a shorter coherence length creating a spatio-temporal speckle pattern which provides some speckle averaging already over a single pulse.\cite{HeinrichNJP08} Furthermore, using a varying speckle pattern across many camera frames, and evaluating the fluctuations, the sectioning capability was recovered, but with a random noise in the image scaling with the inverse root of the number of independent frames, a number limited by signal strength and camera technology. Notably, the image formation relies exclusively on the imaging of the signal onto the camera, and is thus affected by scattering of the sample in the collection path. The absence of a suited technical solution measuring the weak SRL or SRG modulation with an imaging detector presently limits these widefield techniques to CARS and CSRS.

Recently, the wide-field speckle CARS imaging concept was revisited,\cite{FantuzziNPho23} using an optical parametric amplifier providing a much higher repetition rate (515\,kHz). The Stokes speckle pattern was changed only between camera frames, while the pump speckle pattern was changed many times over a single camera frame to provide averaging. A reconstruction algorithm (Wiener filtering) was used to increase the spatial resolution of the retrieved speckle fluctuation image in-plane. Notably, the issue of the speckle noise requiring averging over many frames and the reliance on imaging in the collection path remained.

%Plasmonic nanoparticles to sparsely enhance the CARS signal can be used for parallel detection of signals of many individual NPs, creating a sparse point like sampling.\cite{ZongACSP22} WL: not sure this is fitting here

Instead of wide-field illumination, increasing the number of foci for fast scanning is a known concept in confocal microscopy\cite{OreopoulosMCB14} (Nipkow disk scanners), and also in multiphoton microcopy \cite{BewersdorfOL98,StraubBI98} where camera imaging in the detection path was used to provide the lateral resolution. Sectioning is created by apertures in the detection imaging for confocal microscopy and by the focussed excitation itself in multiphoton microscopy. The inherent trade-off between the number of foci used and the sectioning capability has been investigated.\cite{EgnerJM02} The only multifocal CARS imaging reported\cite{MinamikawaOE09} up top now employed two synchronised Ti:Sapphire oscillators delivering 5\,ps pulses at 80\,MHz repetition rate. A rotating microlens array created an average of seven focal spots moving across the sample at any given time, and CARS was imaged onto a camera, which during the exposure time averaged the signal over the microlens positions fully covering the imaged area. Due to the low pulse energy available, the number of foci and the speed was rather limited, and while an application to cell imaging was shown, \cite{MinamikawaJBO11} the method was not pursued further.

In the present work, we revisit the multi-focal CARS microscopy concept, and demonstrate imaging with a much larger number of 1089 foci at high speed. Importantly we introduce the use of a camera as a multi-segment detector to suppress the detrimental influence of scattering in the detection imaging. The commercial availability of high pulse energy and high repetition rate laser systems such as employed in \Onlinecite{FantuzziNPho23} enables the excitation of a large number of foci without compromising on the CARS efficiency. Furthermore, the availability of fast and low noise CMOS cameras allows read out of the CARS image for every position of the foci. This enables to use the camera images to determine the signal created by each individual focus by summing over a defined region for each focus, emulating the non-imaging single channel detectors employed in single point scanning techniques. The spatial resolution of the resulting CARS image is created by the multiphoton excitation process, as in single point scanning techniques, allowing for deeper penetration into scattering samples.

\section{\label{sec:methods}Methods}

\subsection{Samples\label{subsec:samples}}
Sample O providing a homogeneous reference was prepared by pipetting 10 \textmu L of olive oil (Bio Basso by Basso Fedele \& Figli srl) onto a glass coverslip (\#1.5, $24 \times 50$\,mm$^2$, Thermo Scientific) and was spread across by adding a second equal coverslip for imaging.

Sample A providing bead arrays in water was prepared by pipetting 10\,\textmu L of 1\% solid suspension PS beads of 2\,\textmu m diameter (124, Phosphorex) onto a glass coverslip (\#1.5, $24 \times 50$\,mm$^2$, Thermo Scientific). The mixture was left to dry at ambient conditions and 10\,\textmu L of distilled water was pipetted onto the coverslip and spread across the sample by adding a second equal coverslip for imaging.

Sample B providing a bead mixture in oil was prepared by mixing 10\,\textmu L of 1\% solid suspension of polymethyl methacrylate (PMMA) beads of 20\,\textmu m diameter (MMA20K, Phosphorex) and 10\,\textmu L of 1\% solid suspension polystyrene (PS) beads of 10\,\textmu m diameter (118, Phosphorex) on a glass coverslip as above. The mixture was left to dry under ambient conditions for 30\,mins. Subsequently, 10\,\textmu L of olive oil was pipetted onto the coverslip and spread across the sample by adding a second equal coverslip for imaging.

Sample C providing small beads to demonstrate the point spread function (PSF) was prepared by pipetting 10\,\textmu L of 1\% solid suspension PS beads of 0.5\,\textmu m diameter (108, Phosphorex) onto a coverslip as above and dried at ambient conditions.

Human skin sarcoma samples to demonstrate imaging performance were prepared by fixing them with 10\% formalin and dehydrating it with ethanol. The embedding was performed by using paraffin and the sample was sectioned with a microtome. The sample was stained by using a standard eaosin and hematoxylin staining procedure.\cite{CardiffCSHP14}

\begin{figure*}
	\includegraphics[width=1\textwidth]{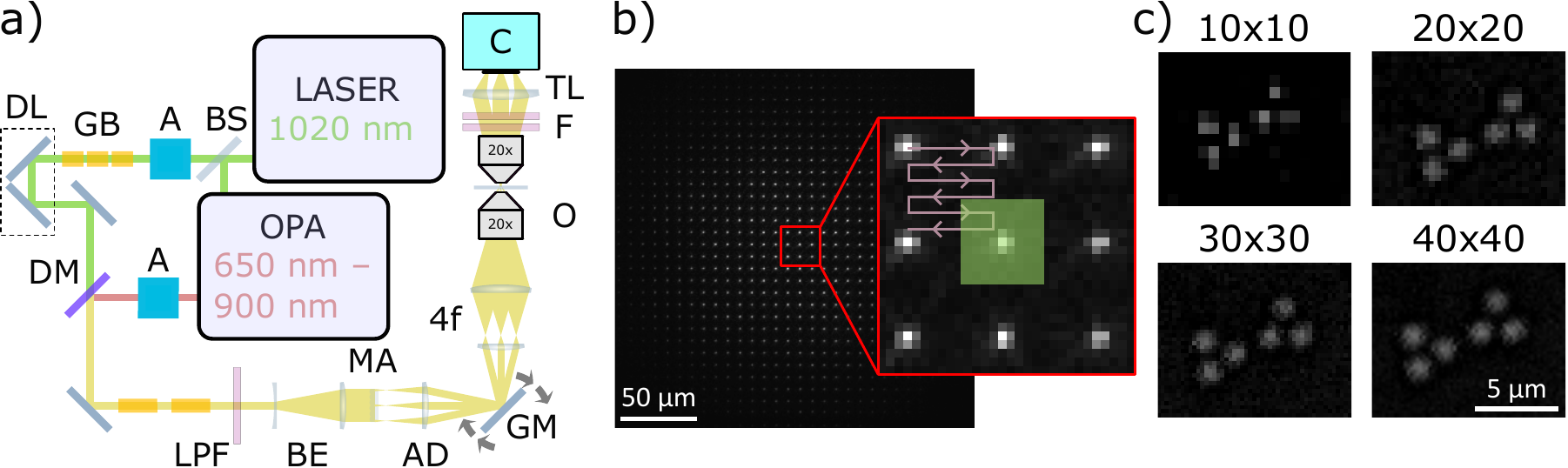}
	\caption{ 
		a) Multifocal CARS experimental setup. BS: beamsplitter; A: power attenuator; GB: glass blocks; DL: delay line; DM: dichroic mirror; LPF: long-pass filter; BE: beam expander; MA: microlens array; AD: achromatic doublet; GM: galvo mirrors; 4f: 4f relay system; O: objective; F: spectral filters; TL: tube lens; C: camera. b) Multifoci raster-scan pattern. At each galvo-mirror position an image of the signals from all focal points is acquired. The pixel values within an square area (green) of $\Si=10$ pixels side length centred at each focal point are summed to provide the focal point signal. $\cgain=20$\,dB, $\texp=4$\,ms, $\nuf=195$\,Hz, grey scale black to white 0 to 1382\,pe. c) Examples of reconstructed images of 2 \textmu m PS beads (sample A) for position grids of $\Ng\times\Ng$ points and \Ng\ of 10, 20, 30, and 40. Grey scale black to white 0 to 1383\,pe.} 
	\label{fig:Setup}
\end{figure*}

\subsection{\label{subsec:setup}Experimental Setup}

The experimental setup used is sketched in \Fig{fig:Setup}a. An amplified laser system (PHAROS, Light Conversion) provides transform-limited 290\,fs pulses of 1020\,nm center wavelength at 200\,kHz repetition rate and 20\,W average power. 5\,W of the beam are used as Stokes beam for CARS, while the remaining power is used for an optical parametric amplifier (OPA) (ORPHEUS-F, Light Conversion) which provides 130\,fs pulses tunable from 650 to 900\,nm  with 1\,W average power, used as pump beam for CARS. The powers of pump and Stokes are controlled by attenuators consisting of a half-wave plate (Castech) and a Glan-Taylor polarizer (GT10-B, Thorlabs). The Stokes beam is transmitted through 60\,mm of SF-66 glass blocks (II-VI incorporated) and an optical delay line (VT-80, PI) with silver retroreflectors (Eksma optics) to control the temporal overlap of pump and Stokes at the sample. Both beams are combined into the same spatial mode by a dichroic mirror (high reflection from 600 to 950\,nm and high transmission from 970 to 1300\,nm, Eksma Optics) and then pass through 145\,mm of SF-57 glass blocks (II-VI incorporated) for spectral focussing.\cite{LangbeinJRS09,GudaviciusJRS24} The pump at 785\,nm accrues 33350\,$\fss$ of group delay dispersion (GDD) stretching it to 723\,fs, while the fixed Stokes accrues 34450\,$\fss$ GDD stretching it to 439\,fs. The difference in GDD can be compensated using a commercially available pulse compressor (Light Conversion) which can add up to 1300\,$\fss$ GDD to the pump beam. Both beams then pass through a long-pass filter (FEL0700, Thorlabs) to remove components below 700\,nm wavelength, spectrally overlapping with the CARS signal (mostly residuals of the white-light generation seed of the OPA). A beam expander (plano-concave lens of focal length $f=-50$\,mm and plano-convex lens of $f=150$\,mm, Casix) increases the beam size from 3\,mm to 9\,mm. The beams then are focussed by a silica microlens square array (15-811, Edmund Optics) of $10\times 10$\, mm$^2$ size, 300\,\textmu m pitch, and $f=8.7$\,mm, creating $33 \times 33$ focal spots. The generated beamlets are collimated by an achromatic doublet (PO-PK-L3, II-VI incorporated) of $f=100$\,mm, and are merging at its back-focal plane which is located at the x-y galvo mirrors (6215H, Cambridge Technology) scanning the beam directions. The beamlets are then relayed with four-times magnification (closely-spaced pair of AC300-100-B and a AC508-200-B-ML, Thorlabs), expanding the beamlets to $13.8 \times 13.8$\,mm$^2$ square size and imaging the galvo mirrors into the 15\,mm diameter back focal plane of the excitation objective (plan apo lambda 20$\times$ 0.75 NA, MRD00205, Nikon), which focuses the beamlets into the sample. The transmitted light is collected by a second objective (plan apo VC 20$\times$ 0.75\,NA, 1501-9398, Nikon) and then passes through a short-pass and a band-pass optical filter (FESH0750 and FB650-40, Thorlabs) to isolate the CARS signal. A camera lens of $f=50$\,mm and $f$/2.8 aperture (NMV-50M23, Navitar) images the sample with a magnification of 5 onto a camera of $2448 \times 2048$ pixels of 3.45\,\textmu m size, $\sigr=2.44$\,e read noise, 10\,ke full well capacity, and USB3.1 interface (BFS-U3-51S5M-C, FLIR). The 18\,mm aperture of the camera objective accommodates the 15\,mm diameter of the collection objective back focal plane and allows for some off-axis beam walk-off. We note that the 5 times magnification onto the camera is slightly below the 7 times required for Nyquist sampling, which reduces the number of pixels to be read per focal spot, improving read noise and speed. The camera is triggered via the software driving the galvos with a frame rate $\nuf$, and the exposure time \texp\ is set to allow this frame rate, which requires $\texp+\tdel<1/\nuf$ with the delay time $\tdel=75$\,\textmu s. The camera is read with 8bit pixel format for highest speed, via a software written in LabView, and the frames are saved as bitmaps and processed off-line. We note that real-time processing of the data would be feasible with more software development. We used a camera analog gain \cgain\ of 10dB, 20dB or 30dB, providing 14.6, 4.61, or 1.46 e/count, respectively, and a digital offset of 5 counts to avoid zero-clipping. The dark signal was subtracted from each frame before further processing.  Powers of excitation beams given later are measured at the sample.

\subsection{Image acquisition \label{subsec:image}}

The CARS signal generated in the sample is imaged onto the camera, showing the $33 \times 33$ grid of beamlet foci, as seen in \Fig{fig:Setup}b for the homogeneous sample O. The intensities of the foci are varying across the array due to the variation of the beam intensity across the lenslet array considering the pump and Stokes excitation beams of Gaussian beam diameter (at $1/e^2$ intensity) of 9\,mm, resulting in a gradual reduction towards the periphery of the array. Notably, CARS is a third-order signal proportional to $\Ip^2\Is$, with the pump intensity \Ip\ and the Stokes intensity \Is, resulting in a diameter of the CARS intensity of about 5\,mm, a factor of $\sqrt{3}$ smaller than the beams. 

The galvo-mirrors are used to shift the foci in a raster-scan pattern with \Ng\ steps in each direction, filling the space between foci, as sketched in \Fig{fig:Setup}b. We note that this could also be achieved by simply laterally shifting the microlens array. An image $P_{kl}$ is acquired at each rasterscan index $(k,l)$ in $x$ and $y$ direction for $k,l=0,1,..,\Ng-1$ . The foci are separated by 7.5\,\textmu m on the sample, which corresponds to $\Sg=10.9$ pixels on the camera. The focal point positions \rnm\ of the beamlets of index $(n,m)$ in $x$ and $y$ directions, $n,m = 0,1,..,32$, are determined using the signal peak positions in the image $P_{00}$ from the homogeneous sample. The camera centering and rotation is aligned so that the central row and column of foci are straight and deviating by less than half a pixel from the horizontal and vertical direction. A second-order polynomial $x(n)=x_0+x_1 n + x_2 n^2$ is then fitted to the horizontal positions of the central row of peaks ($m=16$), and equivalently  $y(m)=y_0+y_1 m + y_2 m^2$ to the vertical positions of the central column of peaks ($n=16$), and the position of any peak is then taken as $\rp(n,m)=(x(n), y(m))$ for image reconstruction. The second-order terms cater for small cushion distortions of the imaging. The nominal focus positions in $P_{kl}$ are then taken as $\rp(n+k\Sg/\Ng,m+l\Sg/\Ng)$ and are providing the signal for the pixel position $(i,j)=(n \Ng+k, m \Ng+l)$ in the excitation image, having a pixel size on the sample of 7.5\,\textmu m/\Ng. 

\begin{figure*}
	\includegraphics[width=1\textwidth]{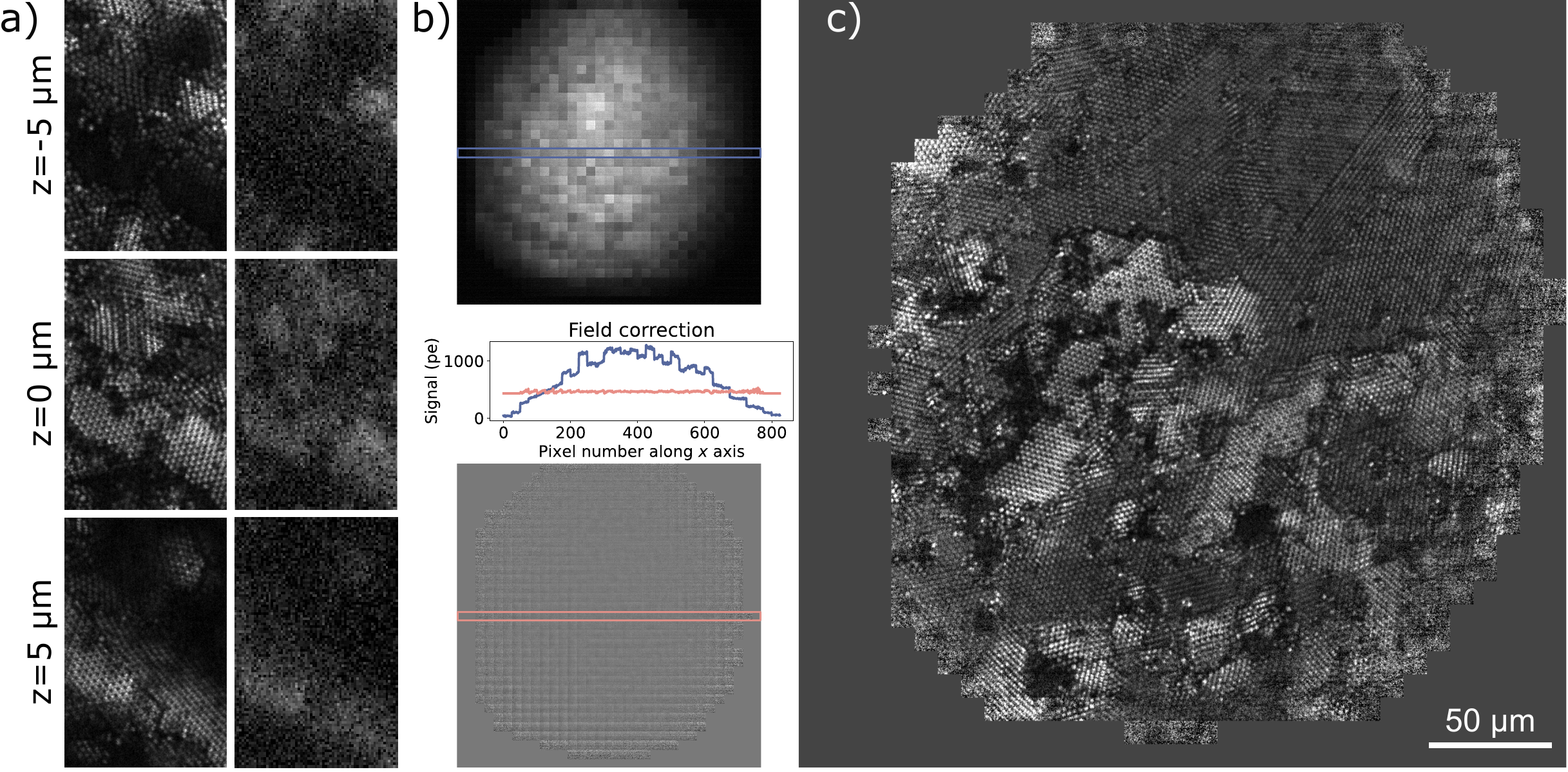}
	\caption{a) Excitation imaging (left) compared to detection imaging (right) at different \textit{z} planes as indicated. CARS of sample A (2\,\textmu m PS beads) at 2950\,\rcm\ with 20\,mW pump power and 50\,mW Stokes power.  $\Ng=30$, $\nuf=195$\,Hz, $\texp=4$\,ms, $\cgain=30$\,dB, $\Si=8$, greyscale 0 to 1399\,pe (left)  and 0 to 19575\,pe (right).  b) Correction of beamlet-dependent signal strength. CARS imaging of the homogeneous sample O (oil) at 2850\,\rcm\ with 20\,mW pump power and 50\,mW Stokes power. Top: image $p_{ij}$ shows a tile structure from imperfections in the microlens array, and vignetting by the excitation beam. $\Ng=25$, greyscale 0 to 1856\,pe. The center cross-section (blue line) is shown below. Applying the flat-field correction (\Eq{eq:corr}), the resulting image \pcij\ is rather uniform (bottom, greyscale 0 to 1114\,pe) as confirmed by the cross-section (red line).  c) Example of a flat-field corrected image \pcij\ of sample A, greyscale 0 to 466\,pe, $v=1/3$, settings as in a).}
	\label{fig:PatternRef}
\end{figure*}

The pixel value $p_{ij}$ of the excitation image is evaluated by summing the pixel values of $P_{kl}$ over a square area with edge size \Si\ centred on the nominal focus positions \rp\ (see  green area in \Fig{fig:Setup}b), noting that avoiding overlap between foci requires $\Si<\Sg$.  The number of pixels summed is $\Si^2$, providing a read noise of $\sigrs=\Si\sigr$, proportional to \Si. The variation of $p_{ij}$ with \Si\ depends on the extension of the imaged signal in $P_{kl}$. In homogeneous samples the extension is below 2 pixels, as seen in \Fig{fig:Setup}b, thus choosing $\Si=3$ seems appropriate to achieve highest signal to noise. However, the extension is affected by sample inhomogeneities in the path from the beamlet focus to the camera, as exemplified in \Fig{SM-fig:PS_WGM}. Thus the immunity of $p_{ij}$ against sample inhomogeneities increases with \Si, and is maximised for $\Si=10$, as shown in \Fig{SM-fig:int_area}. Excitation images of 2\,\textmu m PS beads (sample A) at 2950\,\rcm\ are given in \Fig{fig:Setup}c using different $\Ng$, showing the increasing definition of the imaged beads with \Ng. Nyquist sampling is achieved for $\Ng \approx 30$, providing a pixel size of 250\,nm.

The use of the camera to define a multi-segment detector separating the signal of each focus in the excitation imaging can be compared with the detection imaging employed in previous multifocal approaches,\cite{BewersdorfOL98,MinamikawaOE09} where the image $P$ is given by the sum of $P_{kl}$ over all $(k,l)$. The resulting detection imaging is shown in \Fig{fig:PatternRef}a for sample A in a region with stacked patches of closely packed 2\,\textmu m PS beads in water, imaged at different axial planes. A significant blurring and reduced contrast in detection imaging (right column) compared to excitation imaging (left column) is observed. We note that the detection imaging pixel size of 0.69\,\textmu m is sufficient to resolve the beads. Thus, while the excitation imaging results in a PSF given by the excitation focus as in single point scanning, detection imaging results in blurring due to sample inhomogeneities, an issue intrinsic to the previous multifocal approaches. 

\begin{figure*}
	\includegraphics[width=1\textwidth]{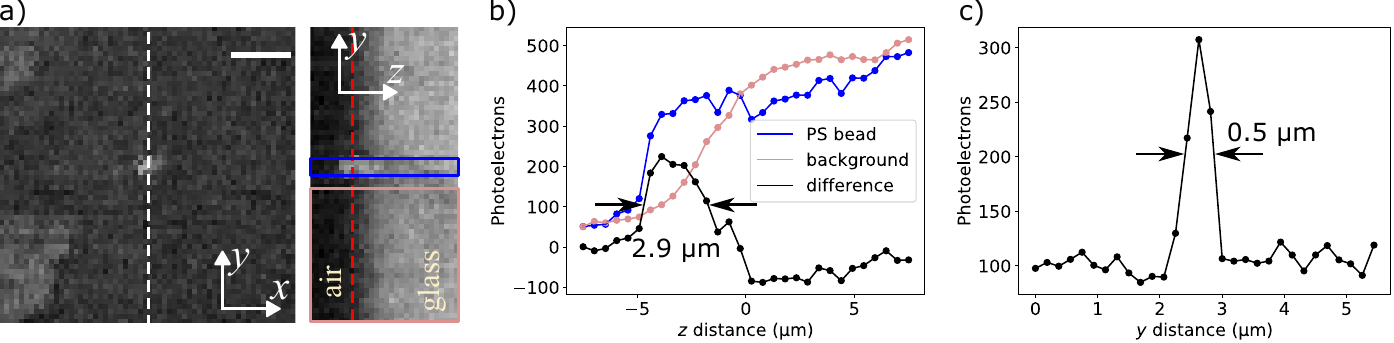}
	\caption{a) CARS PSF measured on a 0.5\,\textmu m PS bead (sample C) using 30 \textit{z}-planes with 0.5\,\textmu m step. Pump 785\,nm, 20\,mW and Stokes 50\,mW, 3050\,\rcm. $\Ng=40$, $\Si=10$, $\cgain=30$\,dB, $\nuf=100$\,Hz. \texp=8\,ms. a) $xy$ and $yz$ sections, with corresponding positions indicated by dashed lines, greyscale 0 to 583\,pe, scale bar 2\,\textmu m. b) \textit{z}-dependent signal of the glass-air interface without PS bead (pink, averaged over the pink rectangle in a), with the PS bead (blue, averaged over blue rectangle in a), and their difference (black), showing the axial PSF. c) signal along white dashed line in a) showing the lateral PSF. } 
	\label{fig:PSF}
\end{figure*}

\subsection{Image post-processing \label{subsec:processing}}

Variations between microlenses lead to variations in the CARS signal from different foci. Moreover, these differences are modulated by sample inhomogeneities. Static, sample independent variations can be corrected using an image of a homogeneous sample and calculating the ratio $r_{nm}$ of the average pixel value over the tile from an individual focus $A_{nm}=\langle p_{(n \Ng+k, m \Ng+l)}\rangle$ to the average value of all tiles $\langle A_{nm} \rangle$. To exclude tiles of low signal, exhibiting excessive noise after correction, tiles with a ratio $r_{nm}$ lower than a threshold value $v$ are set to the average value of the other tiles $\Av=\langle A_{nm} \rangle_{r_{nm}>v}$. The flat-field corrected pixel values are accordingly calculated as
\begin{equation} 
	\pcij=
	\begin{cases}
	p_{ij}/r_{nm} \text{,} &\text{ if }  r_{nm} \geq v\\
	{\Av} \text{,} &\text{ if }  r_{nm} < v
	\end{cases}\,.
\label{eq:corr}
\end{equation} 
The uncorrected pixel values for the homogeneous sample O are presented in \Fig{fig:PatternRef}b. A tile structure from imperfections in the microlens array, and vignetting by the finite excitation beam size are visible. Applying the flat-field correction \Eq{eq:corr}, the resulting image \pcij\ is rather uniform (see \Fig{fig:PatternRef}b bottom and red cross-section). There is a remaining gradient of some $\pm 5$\% across the tiles, which is attributed to clipping of the excitation beams and the CARS signal by the objectives and could be reduced optimising the optical layout. 

A corrected image of sample A with 2\,\textmu m PS beads is presented in \Fig{fig:PatternRef}c, showing the high imaging quality achieved. Notably, even after correction, a tiling pattern is still visible. We attribute this to a distortion of the individual foci by the sample, which can reduce or increase the CARS signal depending on the distortions present without the sample due to the imperfection of the microlens array. Therefore the relative intensity of the foci does depend on the refractive index inhomogeneities of the sample itself, and cannot be fully compensated by a reference measurement on a homogeneous sample. These effects could be mitigated using a higher quality microlens array.

As alternative to hardware improvements, we developed an algorithm to minimise the tile pattern {\it a posteriori}. This algorithm determines correction factors $C_{nm}$ for each beamlet to minimise the step heights at the tile edges over all tiles. To do so, we determine the edge ratios of a tile $nm$, 
\be
u_{nm \rightarrow} =  \frac{\sum_l \tilde{p}_{((n+1)\Ng-1, m \Ng+l)}}{\sum_l \tilde{p}_{((n+1)\Ng, m \Ng+l)}} \\ 
\ee
for the right edge, and 
\be
u_{nm \downarrow} =  \frac{\sum_k \tilde{p}_{(n\Ng+k, (m+1)\Ng-1)}}{\sum_k \tilde{p}_{(n\Ng+k, (m+1)\Ng)}} 
\ee
for the bottom edge. These edge ratios are modified by applying the tiling correction factors $C_{nm}$, and we find the factors minimising the deviation of the modified ratios from unity using
\bea \label{eqn:tiling}
	C_{nm} &=& \mathrm{argmin} \sum_{n,m} \\ \nonumber && {\left|\log\left(\frac{u_{nm \rightarrow} C_{nm}}{C_{n+1,m}}\right)\right| + \left|\log\left(\frac{u_{nm \downarrow} C_{nm}}{C_{n,m+1}}\right)\right|}\,.
\eea
The tiling corrected data is then given by $\pctij=\pcij C_{nm}$. An example of the tiling correction is discussed later in \Fig{fig:Beads}.

\begin{figure*}
	\includegraphics[width=1\textwidth]{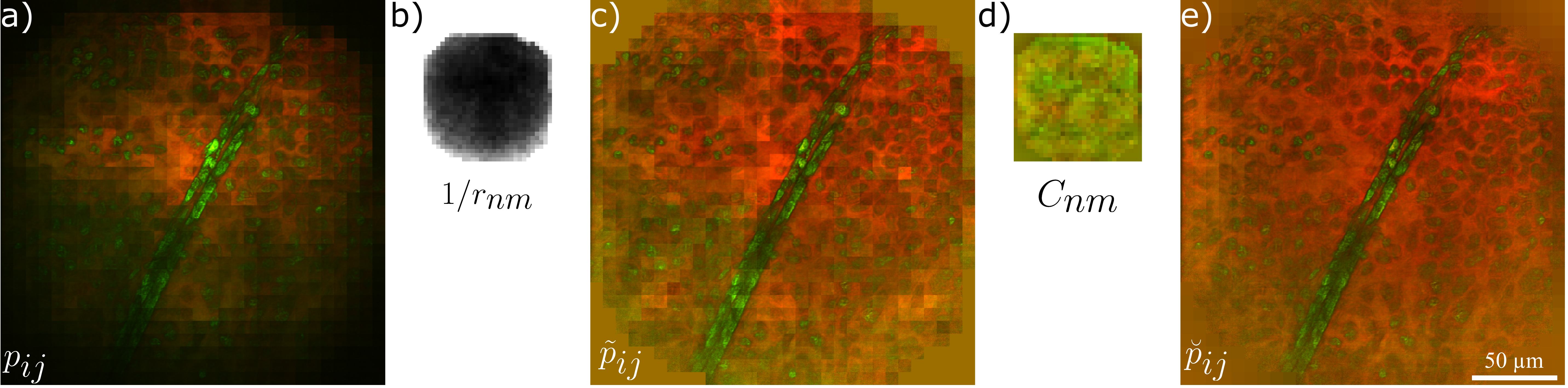}
	\caption{CARS false colour RGB image of a human skin sample demonstrating the flat field and  tiling correction. $\nuf=50$\,Hz, $\texp=15$\,ms, $\cgain=20$\,dB, $\Ng=30$, $\Si=10$. Pump 785\,nm, 18\,mW; Stokes 48\,mW.  CARS was imaged at 2935\,\rcm\ (red) and then 2970\,\rcm\ (green) by changing the pump-Stokes delay.	a) Uncorrected $p_{ij}$, range $0-10138$\,pe for 2935\,\rcm\ and $0-5530$\,pe for 2970\,\rcm. b) Flat-field correction factors $1/r_{nm}$ using the glass coverslip of the sample as homogeneous reference, greyscale from 0 to 4. c) Flat-field corrected data \pcij, range 0 to 3226\,pe for 2935\,\rcm\ and 0 to 5529\,pe for 2970\,\rcm. d) Tiling correction factors $C_{nm}$ with color coding as c) on a range from 0 to 2. e) Tiling corrected \pctij, same range as in c). }
	\label{fig:Pattern}
\end{figure*}

\section{Results}

\subsection{Point spread function \label{subsec:psf}}
\noindent To measure the experimental three-dimensional PSF of the setup, a 0.5\,\textmu m diameter PS bead on glass in air (sample C) was imaged. Flat-field corrected images \pcij\ in the \textit{x-y} as well as \textit{y-z} planes are presented in \Fig{fig:PSF}. The in-plane PSF full width at half maximum (FWHM) is 0.5\,\textmu m, while the axial PSF FWHM is 2.9\,\textmu m, comparable to previous single point scanning data using an 0.75\,NA objective.\cite{KarunaJRS16}

\subsection{Tiling correction \label{subsec:tiles}}

To demonstrate the efficacy of the tiling correction on biomedical samples of interest, a human skin sample (see \Sec{subsec:samples}) was imaged at 2935\,\rcm\ and 2970\,\rcm. The resulting data $p_{ij}$ using $\Si=10$ is presented in \Fig{fig:Pattern}a. The vignetting due to the excitation beam profiles is clearly visible, together with the tiling. After flat-field correction using $r_{nm}$ from sample O and $v=1/3$ (see \Fig{fig:PatternRef}), shown in \Fig{fig:Pattern}b, the resulting \pcij\ given in \Fig{fig:Pattern}c shows that the vignetting has been corrected, but the tiling, while somewhat reduced, is still clearly visible. The change in the tile pattern indicates that the origin of the remaining pattern is different from the tiling in a homogeneous medium. Applying the tiling correction \Eq{eqn:tiling} yields the factors $C_{nm}$ shown in \Fig{fig:Pattern}d. The corrected \pctij\ given in \Fig{fig:Pattern}e is void of significant tiling, confirming that the correction is effective. The visible structure is attributed to basal cells (circular objects) in a collagen matrix and elastin (tubular shape).

\begin{figure*}
	\includegraphics[width=1\textwidth]{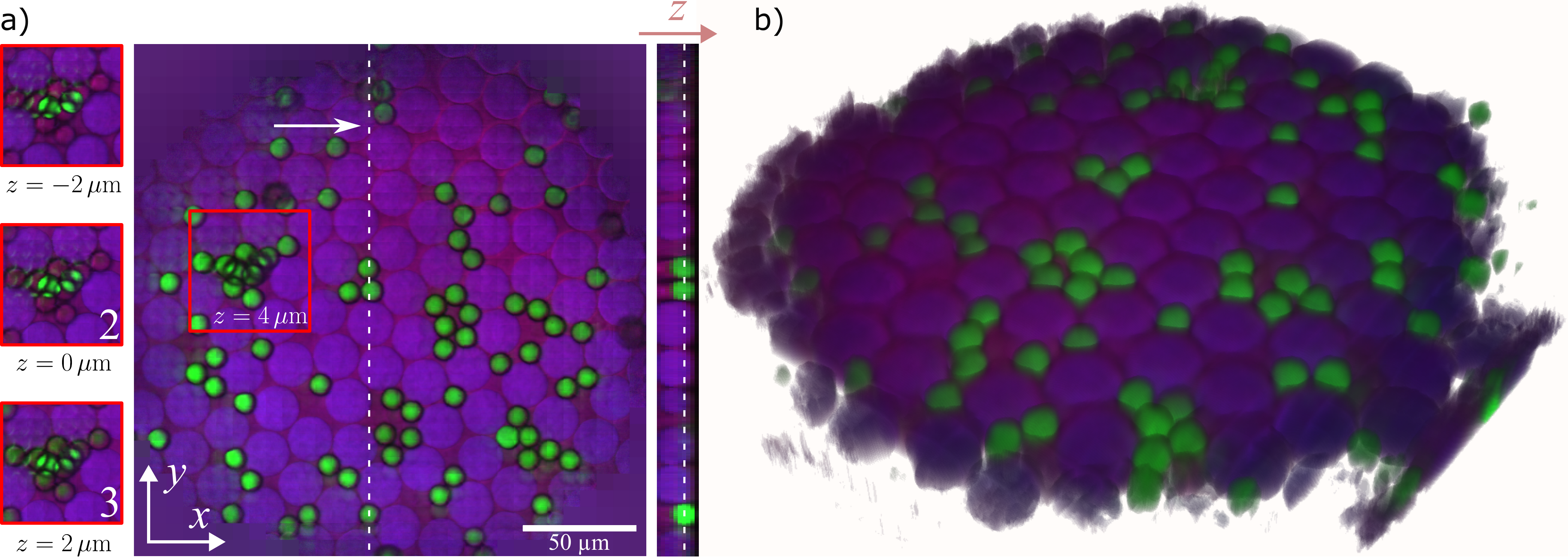}
	\caption{Three-dimensional CARS imaging of sample B at 3050\,\rcm\ (green) highlighting the 10\,\textmu m PS beads, at 2950\,\rcm\ (blue) highlighting the 20\,\textmu m PMMA beads, and at 2930\,\rcm (red) showing the oil used as embedding medium. $\nuf=100$\,Hz, $\texp=5$\,ms, $\cgain=10$\,dB, $\Ng=30$, $\Si=10$. Pump 785\,nm, 10\,mW; Stokes 50\,mW. Stack of 10 \textit{z}-steps of 2\,\textmu m size. a) Images \pctij\ after flat-field ($v=1/5$) and tiling correction using value ranges of 0 to 14600\,pe for 2930\,\rcm, 0 to 17520\,pe for 3050\,\rcm, and 0 to 29142\,pe for 2950\,\rcm. a) $x-y$ section, with a zoom of an area indicated as red square at different \textit{z} planes with 2\,\textmu m increments shown on the left, and a $y-z$ section shown on the right. The white dashed lines indicate the $x$ and $z$ position of the $y-z$ and $x-y$ sections. b) 3D reconstruction of \pctij.} 
	\label{fig:Beads}
\end{figure*}

\subsection{Three-dimensional bead imaging \label{subsec:beads}}

To demonstrate the \textit{z}-sectioning capability on strongly scattering structures, sample B hosting a mixture of 20\,\textmu m PMMA beads and 10\,\textmu m PS beads in oil was imaged. A $z$-stack over 10 positions with 2\,\textmu m increments was measured. Using different pump-Stokes delays, CARS stacks at 2930\,\rcm\ dominated by oil, at 2950\,\rcm\ dominated by PMMA, and at 3050\,\rcm\ dominated by PS were sequentially recorded. Flat-field and tiling correction was applied, and the resulting \pctij\ is shown in \Fig{fig:Beads}. We can see that the method is able to create well resolved $z$-sectioning also for strongly scattering samples. Comparing this with detection imaging as previously used\cite{BewersdorfOL98,MinamikawaOE09} and shown in \Fig{SM-fig:img_methods}, the significant advantage of excitation imaging created by the non-linear exitation PSF is evident. Nevertheless, the 10\,\textmu m PS beads with their large refractive index contrast to olive oil (1.59 to 1.46) provide such large deviations in the detection imaging that the signals from different foci are mixed, as detailed in \Fig{SM-fig:PS_WGM}. This mixing leads to the shadowing of the lower PS beads by the beads lying on top of them in \Fig{fig:Beads}a 1-3. Considering that the shadowing shows the obstacles in detection direction, increasing the NA of excitation and detection would spatially blur and thus reduce the shadowing.

\subsection{CARS and SHG imaging of human skin \label{subsec:skin}}

\begin{figure}[b]
	\includegraphics[width=\columnwidth]{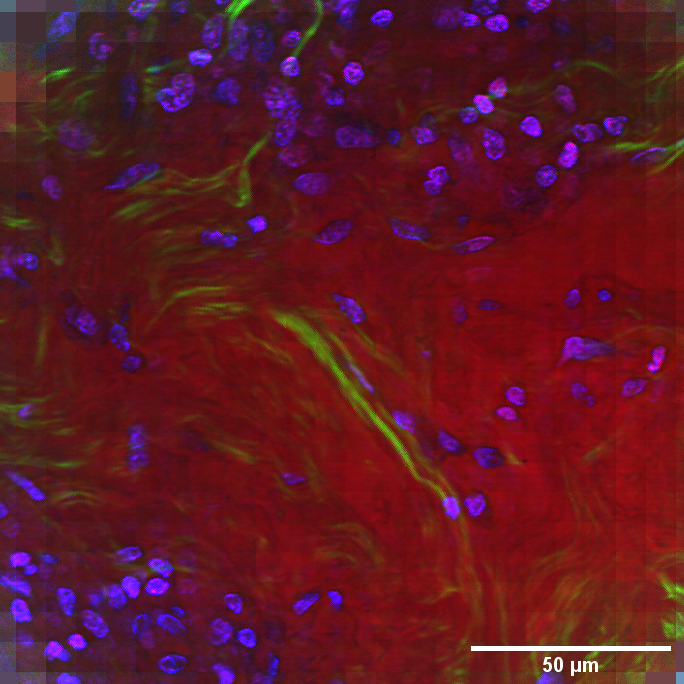}
	\caption{CARS and SHG imaging of human skin tissue. $\nuf=200$\,Hz, $\texp=4$\,ms, $\cgain=20$\,dB, $\Ng=30$, $\Si=10$. Pump 785\,nm 20\,mW, Stokes 50\,mW. Flat-field ($v=1/5$) and tiling correction. Blue channel is CARS at 2970\,\rcm\, red channel is CARS at 2935\,\rcm, and green channel is SHG. Value ranges are 0 to 8294\,pe for the red, 0 to 4608\,pe for the green, and 0 to 3686\,pe for the blue channel.} 
	\label{fig:CARS_SHG}
\end{figure}

To demonstrate the performance of the multifocal approach on medically relevant scattering samples, a human skin tissue was imaged (see \Sec{subsec:samples}). CARS was recorded at 2935\,\rcm\ and 2970\,\rcm. To measure SHG of the Stokes, the FB650-40 filter was replaced by a BP520-40 (Thorlabs), and the pump was blocked. The glass blocks were not removed for convenience, but we note that the SHG signal would be increased about 4 times by doing so. The resulting image encoding all three data as false colour is presented in \Fig{fig:CARS_SHG}. We attribute the structure to basal cells (blue), elastin fibres (green), and collagen matrix (red).
	
\section{\label{sec:conclusions}Conclusions}
In this work we have demonstrated multifocal CARS microscopy with 1089 foci, enabled by a high repetition rate amplified oscillator and optical parametric amplifier.
Importantly, different from previous multifocal approaches,\cite{BewersdorfOL98, MinamikawaOE09} we introduced employing the camera detection only to separate the signals of the different foci, and not for imaging. This retains the insensitivity to sample scattering afforded by the non-linear excitation PSF created by longer wavelength light, which is the hallmark of point-scanning CRS. Notably, speckle widefield CARS\cite{HeinrichNJP08,FantuzziNPho23} is using detection imaging and also has intrinsic speckle noise. 

The speed of the technique in the present setup is limited by the camera readout, while the laser source caters for up to ten times higher excitation powers, and correspondingly 1000 times higher CARS powers, allowing to increase the speed accordingly, up to $10^5$ frames per second. This would enable a 100\,Hz frame rate for megapixel CARS images, corresponding to 10\,ns pixel rate, about an order of magnitude faster than the fastest single point scanning systems. Notably, due to the 1000 foci, this is achievable with similar single focus power densities using a 100 times lower repetition rate than single point systems. Since the camera is only used to separate the different foci rather than to image the sample at the Nyquist limit, only a small number of pixels are required, for example $100\times100$ pixels would suffice for 1000 foci, which can be read by present camera technology at up to $10^6$ frames per second, opening the perspective to megapixel CARS imaging with 1000\,Hz frame rates.

As future improvement of the method, time-multiplexing\cite{AndresenOL01} is a promising approach not only for suppression of cross-talk between foci as demonstrated for TPF,\cite{AndresenOL01} but also to address different vibrational frequencies via spectral focussing, allowing for a spectrally resolved imaging where different foci address different vibrational resonances, doing away with the sequential spectral tuning.

\begin{acknowledgments}
We wish to acknowledge Jonas Berzinš (Light Conversion) for proofreading and commenting on the draft, Skaidra Valiukevičienė (Lithuanian University of Health Sciences) for human skin sarcoma samples, and Paola Borri for support. This project has received funding from the European Union’s Horizon 2020 research and innovation programme under the Marie Sk\l odowska-Curie grant agreement No 812992.
\end{acknowledgments}

\appendix

\section{\label{sec:int_area} Influence of detection region size on excitation image}

\begin{figure*}
	\includegraphics[width=1\textwidth]{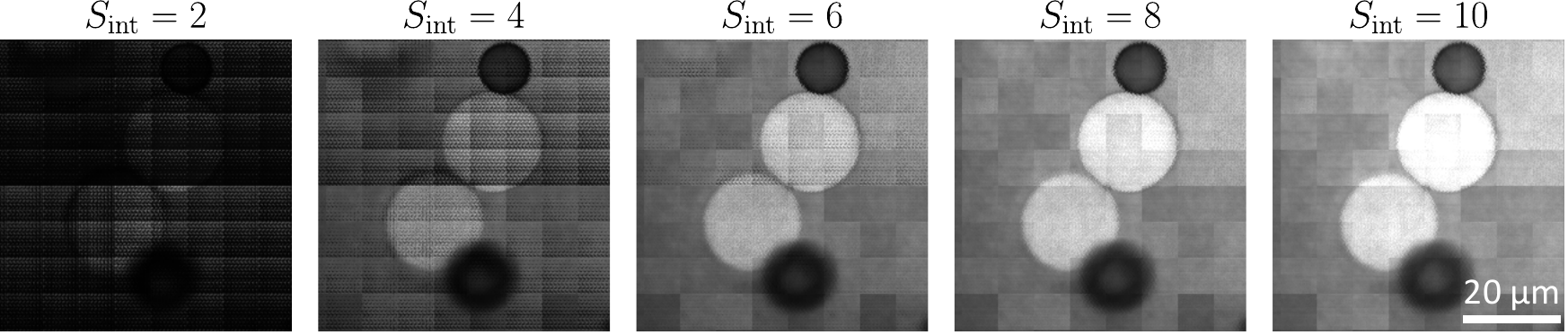}
	\caption{Dependence of excitation image $p_{ij}$ on the detection region size \Si\ for sample B  in a region showing two 10\,\textmu m PS and two 20\,\textmu m PMMA beads embeded in oil at 2950\,\rcm.  $\nuf=50$\,Hz, $\texp=15$\,ms, $\cgain=20$\,dB, $\Ng=30$, pump 785\,nm, 10\,mW; Stokes 50\,mW. Results for $\Si$ from 2 to 10 are shown as labelled. Greyscale range black to white 0 to 6480\,pe.
	}
	\label{SM-fig:int_area}
\end{figure*}

The size \Si\ of the region used to sum the signal for each excitation focus is determining the properties of the resulting exictation image $p_{ij}$, as discussed in \Sec{subsec:image}. This dependence is exemplified in \Fig{SM-fig:int_area}, showing $p_{ij}$ for detection region sizes \Si\ from 2 to 10 for PS and PMMA beads embeded in oil. One finds that for small $\Si$, for which the detection region does not fully cover the signal from each focus, stripes appear due to the discrete positioning governed by the pixelation of $P$. With increasing \Si, the fraction of the collected signal increases and saturates, providing an image free of the pixelation artifact, only carrying the tile pattern from the variation between foci. 

\section{\label{sec:img_methods} Detection and excitation imaging of beads}

\begin{figure}
	\includegraphics[width=0.5\textwidth]{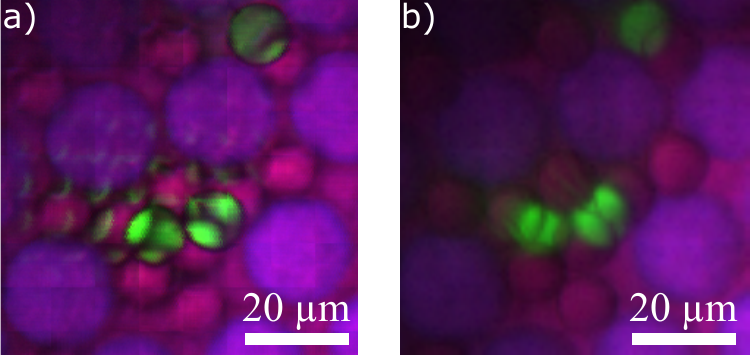}
	\caption{Comparison between excitation and detection imaging for the CARS data shown in \Fig{fig:Beads} of sample B with 10\,\textmu m PS beads and 20\,\textmu m PMMA beads, $z=-2$\,\textmu m. a) Excitation image \pctij\  after flat-field and tiling correction using value ranges of 0 to 21857\,pe for 2930\,\rcm (red), 0 to 23314\,pe for 3050\,\rcm (green), and 0 to 29142\,pe for 2950\,\rcm (blue). b) Detection image $P$, using value ranges of 0 to 197100\,pe for red, 0 to 131400\,pe for green, and 0 to 393300\,pe for blue.
	}
	\label{SM-fig:img_methods}
\end{figure}

\Fig{SM-fig:img_methods} shows a comparison between detection imaging and excitation imaging for the bead sample B. The detection image $P$ is of lower sharpness and contrast compared to the excitation image \pctij. Notably since $\Ng=30$, there are 900 pixels per focus in \pctij, while there are $\Sg^2=119$ camera pixels over the same area, resulting in a higher signal per pixel in $P$.  

\begin{figure*}
	\includegraphics[width=1\textwidth]{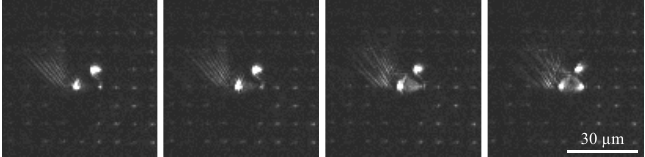}
	\caption{Selected frames $P_{kl}$ of CARS imaging of sample B (data from \Fig{fig:Beads}a1) containing 20\,\textmu m PMMA and 10\,\textmu m PS beads showing strong signal scattering for some foci. From left to right $(k,l)$ is $(1,6)$, $(1,29)$, $(3,26)$, and $(4,2)$. Greyscale 0 to 437\,pe, \cgain=10\,dB, $\texp=5$\,ms, $\nuf=100$\,Hz, $\Ng=30$. Pump 785\,nm, 20\,mW, Stokes 50\,mW, addressing the PS resonance at 3050\,\rcm. All $\Ng\times\Ng=900$ frames acquired are available as movie in the online material (Scattering.avi).
	}
	\label{SM-fig:PS_WGM}
\end{figure*}

The beads in sample B can lead to a strong scattering of the CARS signal in detection imaging, due to the high index contrast beween PS and oil. This is exemplified in selected frames $P_{kl}$ shown in \Fig{SM-fig:PS_WGM}, and in the movie of all $\Ng\times\Ng=900$ frames provided in the file {Scattering.avi}.
The beads can form whispering gallery modes by total internal reflection, leading to the observed standing wave patterns. Notably, the signal is distributed widely across the frame, well beyond the detection region size \Si, resulting in cross-talk between foci creating weak "phantom" images of the PS beads, and inhomogeneous signal within the PS beads, visible in \Fig{SM-fig:img_methods}a. 

%\nocite{*}
%\newpage
%\bibliography{MFCARS,DualCARS,langsrv}% Produces the bibliography via BibTeX.
%aipnum4-2.bst 2019-01-14 (MD) hand-edited version of apsrev4-1.bst
\providecommand{\noopsort}[1]{}\providecommand{\singleletter}[1]{#1}%\providecommand{\noopsort}[1]{}\providecommand{\singleletter}[1]{#1}%
\end{document}